\documentclass[aps,prd,onecolumn,amssymb,eqsecnum,nofootinbib]{revtex4}

\usepackage{epsf}  
\usepackage{graphicx}

\begin{document}


%


%



\title{Two-timescale adiabatic expansion of a scalar field model}

\author{Yasushi Mino}
\email{mino@tapir.caltech.edu}

\affiliation{mail code 130-33, California Institute of Technology, 
Pasadena, CA 91125}

\author{Richard Price}
\email{rprice@phys.utb.edu}

\affiliation{Center for Gravitational Wave Astronomy, 
80 Fort Brown, Brownsville, TX. 78520}

\begin{abstract}
The analysis of gravitational wave data may require greater accuracy
than is afforded by the adiabatic approximation to the trajectory of
and field produced by a particle moving in curved spacetime. Higher
accuracy is available with a two-timescale approach using as an
expansion parameter the ratio of orbital time to radiation reaction
time.  To avoid apparent divergences at large distances, the details
of the method are important, especially the choice of the foliation,
the spacetime surfaces on which the orbital elements are taken to be
constant.  Here we apply the two-timescale approach to a simple linear
model to demonstrate the details of the method. In particular we use
it to show that a null foliation avoids large-distance divergences in
the first-order post-adiabatic approximation, and we argue that this
will be true more generally for a null foliation.
\end{abstract}

\maketitle

\section{Introduction}

A major target of gravitational wave detectors, such as LIGO, but
especially LISA\cite{GW}, is radiation 
from extreme-mass-ratio inspirals (hereafter EMRIs), compact objects,
approximated as point particles, orbiting in the field of much larger
gravitating centers, typically Kerr black holes.  In the zeroth
approximation these particles move through the Kerr background on geodesics
characterized by orbital constants $E$ (energy), $L$ (angular
momentum), and $K$ (Carter constant), and the gravitational radiation
generated by the particle is found with perturbation theory to first
order in the EMRI ratio, the ratio of the particle mass to the
background mass.

In reality, of course, the orbits cannot be geodesics. The loss of
energy and angular momentum to radiation requires that the orbits
evolve. The evolution can be understood either as a  reaction
to the radiation, or as the result of a self force, the force on the
particle due to the spacetime perturbations created by the particle
itself.  In either viewpoint, the deviation from a geodesic is
proportional to the EMRI mass ratio, and for most EMRI ratios means
that the deviation is small enough to justify what is called the
adiabatic approximation: the particle is considered to be moving, at
any moment, on a geodesic with orbital ``constants'' $E,L,K$ that are
functions of time as the orbit evolves. More important, the same
approximation is made for the gravitational waveform generated by the
moving particle: the waveform is to be that from a geodesic orbit
characterized by $E,L,K$, with the three constants varying in time.

In this adiabatic viewpoint the understanding of the orbit and of the
waves it generates is shifted to the question of the evolution of the
orbital constants.  The earliest method for doing this was to balance
the loss of $E$ and $L$ to the radiation energy and angular momentum
going out to infinity and down the horizon\cite{balance}. This method,
though physically appealing, is not well justified except for circular
orbits. Furthermore, it cannot be used to find the evolution of the
Carter constant.  The alternative, a calculation of the self-force
driving the evolution, suffers both from the need for regularization
of the divergent perturbation fields at the particle, and the gauge
aspects of the self force. Some consistency of the balance and the
self force methods has been found, when time averaging of the self
force is used\cite{YM_PRD67}, although in a restricted gauge.

Many technical details remain unresolved. In particular, the
self-force can appear to be entirely gauge dependent on a short
timescale\cite{YM_PrgThPhys113}. Indeed, along with the issue of gauge
invariance, the most pressing problem in this area is the question of
the timescale on which the adiabatic approximation can be made to
work. For gravitational wave analysis, what is needed is a description
that applies over a large change in the orbital parameters. An outline
of such a method has been proposed based on considerations of second
order particle perturbation theory\cite{mod}.

In order to formulate a justifiable scheme for an adiabatic
approximation, a two-timescale analysis 
will be needed in which the adiabatic approximation is only the first
step in an expansion of fields in orders of the ratio of a slow to a
fast timescale.  The mathematical framework for doing this was already
inherent in Ref.~\cite{adi}, in which it was shown  that a
consistent two-timescale approach involves several difficulties.
First, the post-adiabatic effect is at the same order as nonlinear
effects, and the nonlinear gravitational perturbations can result in
both infrared-type and ultraviolet-type divergences that are not easy
to regularize.  Second, there is a weak violation of the gauge
condition for the linear perturbations due to radiation reaction.  The
complications of those issues can obscure technical issues inherent in
the basic idea of the two-timescale method itself. In particular, if
not carried out in the optimal way the two timescale method can lead
to apparent divergences.  This is motivation for 
what is done in the present paper, the application of the
two-timescale approach to a well-understood model: a linear scalar
field coupled to a point particle.

We use Minkowski coordinates $\{t,x,y,z\}$ and we denote the spacetime
coordinates collectively by Greek indices as $x^\alpha
(\alpha=t,x,y,z)$, and we use the Einstein summation rule with the
flat metric, $\eta_{\alpha\beta} = diag(-1,1,1,1)$.  Where appropriate
we switch to spherical spatial $\{r,\theta,\phi\}$ coordinates.
When we refer specifically to the coordinate location of the particle
we use the tilde, as in $\tilde x^\alpha$. 
We denote the spatial coordinates 
by Roman indices as in $x^i (i=x,y,z)$.  
We adopt geometrized units in which $c=G=1$, 
so that mass, length and time have the same dimension.

As a simplest toy model for the practical demonstration in this paper,
we calculate the scalar field $\Phi$ induced by a point charge $q$
according to the 
flat spacetime wave equation,
\begin{eqnarray}
\Box \Phi(x) = \rho(x) \,, \quad 
\rho(x) = q \int d\tau \delta^{(4)}\left(x-\tilde x(\tau)\right) 
\,, \label{eq:feq} 
\end{eqnarray}
where we use the box-operator as 
$\Box = -(\partial_t)^2+(\partial_x)^2
+(\partial_y)^2+(\partial_z)^2$. Here 
$\delta^{(4)}(x)$ is the $4$-dimensional Dirac's delta 
and $\tau$ is the proper time of the orbit. 

For simplicity, we assume that the particle motion follows 
a Newtonian second law
for the potential of a central gravitating mass $M$,
\begin{eqnarray}
{d \over dt}v^i &=& -{M \tilde x^i \over \tilde r^3} +a^i 
\,, \label{eq:eom}
\end{eqnarray}
where $v^i=d \tilde x^i/d \tilde t$ is the ordinary spatial velocity
(coordinate change per coordinate time) and $a^i$ is the ordinary
accerelation (in terms of coordinate time) due to scalar radiation
reaction.

One of the challenges of point particle perturbation theory is to extract 
a finite $a^i$ from the scalar field 
which is divergent along the orbit \cite{sf}.
But our main interest in this paper is the calculation of the first 
post-adiabatic term in a two-timescale expansion. 
For that we shall not need an explicit form of $a^i$, 
and will treat it as a known quantity.
As one additional simplification, we will consider in this paper only
the case of quasi-circular orbits (orbits which are circular except
for radiation reaction).

The structure of this paper is as follows. In Sec.~\ref{sec:orb}, 
we calculate the orbital equation 
including the radiation reaction effect. 
In Sec.~\ref{sec:adi0}, we give the adiabatic approximation for the scalar field,  
the leading order of the adiabatic expansion. 
Though the result here is familiar, 
the approach in this section clarifies the geometric meaning
of that approximation and, in particular, introduces the idea 
of the spacetime foliation function for evolving orbital elements.
In Sec.~\ref{sec:adi1}, we calculate 
the first post-adiabatic correction of the scalar field. 
We conclude and summarize in Sec.~\ref{sec:sum}, 
and discuss applications of this new expansion scheme 
to other problems in relativity. In order to have the main text
focus on the central ideas, we have relegated many of the details to 
a set of appendices.

\section{Orbital Evolution with Radiation Reaction}
\label{sec:orb}

In the adiabatic expansion orbits are characterized by the orbital
elements, the integral constants of the orbital equations in the
absence of radiation reaction.  The evolution of the orbit due to
radiation reaction is then described by the evolution of the orbital
elements\cite{osculating}.  
This means that the orbital elements become functions of
``time,'' i.e., of some parameter $f$ along the orbit.  It is through
this dependence that the effects of radiation reaction will appear in
the orbits.  One of the central issues of this paper is to consider
just what time slicing of spacetime is appropriate to the evolution of
the orbital elements.  That is, how is the orbital parameter $f$ to be
promoted to a function of spacetime location $f(x^\alpha)$?

We align our coordinates so that the quasi-circular orbit is 
at $\tilde{z}=0$ (Cartesians) or $\tilde \theta=\pi/2$ (sphericals).
In order to calculate the orbital evolution (\ref{eq:eom}) ,
including the radiation reaction effect, 
it is convenient to use the orbital energy $e$ 
and the $z$-component of the angular momentum $\ell$
defined by 
\begin{eqnarray}
e = {1 \over 2} v^i v^i - {M \over \tilde r} \,, \quad 
\ell = \tilde x v^y - \tilde y v^x \,. 
\end{eqnarray}
These are constants of motion 
which are conserved along the orbit 
in the absence of  radiation reaction. 
Due to the radiation reaction, $e$ and $\ell$ are not constant 
and the evolution equations are 
\begin{eqnarray}
{d \over df}e := {d\tilde t \over df} a^e 
= {d\tilde t \over df}v^i a^i \,, \quad 
{d \over df}\ell := {d\tilde t \over df}a^\ell 
= {d\tilde t \over df}(\tilde x a^y -\tilde y a^x) 
\,. \label{eq:el-ev} 
\end{eqnarray}
Here we have computed the dynamics of the orbits 
with coordinate time $\tilde{t}$, 
but have used $d\tilde{t}/df$, evaluated at the orbit, 
to infer the influence of the dynamics on the orbital elements. 
We have also defined two components of radiation-reaction 
driven acceleration: $a^e$ and $a^\ell$.

Because radiation reaction effect is weak, 
we first integrate the orbital equation of motion 
in the absence of the radiation reaction. 
The orbital coordinates are written as 
\begin{eqnarray}
\tilde r(f) = r_C \,, \quad 
\tilde \theta(f) = \pi/2 \,, \quad 
\tilde \phi(f) = \Omega \left[\,\tilde t(f)-\tilde t(f\!=\!0)\,\right]
+\phi_C 
\,, \label{eq:orb} 
\end{eqnarray}
where $r_C$ and $\Omega$ are the orbital {\it principal} elements, 
and $\phi_C$ is the orbital {\it positional} elements. 
The function $\tilde t(f)$ is obtained 
from the definition of the foliation function, 
$f\left(\tilde t,r_C,\pi/2,\tilde \phi(f)\right)=f$.  
Since the orbit (in the absence of radiation reaction) 
is circular, $e$ and $\ell$ are related by
\begin{eqnarray}
\ell = {M \over \sqrt{-2e}} \,. \label{eq:el-cir}
\end{eqnarray}
It has been shown 
that this circularity relationship is not only correct 
in the absence of radiation reaction, but holds to all orders 
in an adiabatic expansion, 
as long as the radiation reaction is small \cite{cir}.
From this it follows 
that the two components of the radiation reaction acceleration 
always obey
\begin{equation}
a^\ell=-\frac{1}{2}\,\frac{\ell}{e}\,a^e\,.
\end{equation}

For the circular orbit all kinematical quantities are related. 
It is convenient to express them all 
in terms of the central mass $M$ and the orbital speed $v$: 
\begin{eqnarray}
e = -{1 \over 2}v^2 \,, \quad 
\ell = {M \over v} \,, \quad 
r_C = {M \over v^2} \,, \quad 
\Omega = {v^3 \over M} \,. \label{eq:orbx} 
\end{eqnarray}
With these relationships we can regard $v$ as an orbital 
principal element that governs all other principal elements.
The evolution of $v$ itself is 
\begin{eqnarray}
{d \over df} v 
= -{1 \over v} {d\tilde t \over df} a^e 
= -{v^2 \over M} {d\tilde t \over df}a^\ell \,. 
\end{eqnarray}
For the relationships in (\ref{eq:orbx}) to hold in the 
presence of radiation reaction, $\Omega$ must be defined 
to be $\Omega:=d\tilde{\phi}/d\tilde{t}$.
According to (\ref{eq:orb}), 
this requires that $\phi_C$ evolves according to 
\begin{eqnarray}
{d \over df}\phi_C 
&=& -\left[\,\tilde t(f)-\tilde t(f\!=\!0)\,\right] {d \over df}\Omega 
\,. \label{eq:phi}
\end{eqnarray}

Because the evolution of the orbital elements is derived perturbatively 
{\em after} we calculate the scalar field by the adiabatic expansion, 
we start with the orbit as a function of the orbital elements, 
and we write the source term of the scalar field equation 
as $\rho(x^\alpha) = \rho\left(x^\alpha | C(f(x^\alpha))\right)$ 
where we denote by $C$ the collection of orbital elements.

\section{Adiabatic Expansion - leading order} 
\label{sec:adi0} 

The adiabatic expansion can be elegantly formulated as a two-timescale
expansion: One scale is the dynamical scale of the system denoted by
$T_{\rm dyn}$, such as the orbital period or the wavelength of the
scalar field.  The other is the radiation reaction scale denoted by
$T_{\rm rad}$, such as the timescale for a change in orbital energy
due to radiation reaction.  We assume that the radiation reaction
scale is much longer than the dynamical scale and we use the
dimensionless small value $\mu := T_{\rm dyn}/T_{\rm rad}\ll1$ as
an expansion parameter. That is, we assume that the orbital
principal elements, the elements directly related to $v$,
evolve slowly on the timescale of the radiation reaction. 
To clarify the order of the two-timescale expansion, 
we can use $\mu$ as  the expansion index 
and can replace the $f$-derivatives of $v$ by
\begin{eqnarray}
\left( {d \over df}\right)^n v \rightarrow 
\mu ^n \left( {d \over df}\right)^n v 
\,. \label{eq:exp} 
\end{eqnarray}

From (\ref{eq:orbx}), we have that 
\begin{eqnarray}
{d \over df} r_C \sim {d\over df}\Omega \sim O(\mu) 
\,, \quad
{d^2 \over df^2} r_C \sim {d^2\over df^2}\Omega \sim O(\mu^2) 
\,, \label{eq:exp1} 
\end{eqnarray}
and, from (\ref{eq:phi}), that 
\begin{eqnarray}
{d \over df} \phi_C 
= -\left[\,\tilde t(f)-\tilde t(f\!=\!0)\,\right]
{d \over df}\Omega 
\sim O(\mu) 
\,, \quad 
{d^2 \over df^2} \phi_C 
= -{d\tilde t \over df} {d \over df}\Omega 
-\left[\,\tilde t(f)-\tilde t(f\!=\!0)\,\right]
{d^2 \over df^2}\Omega 
\sim O(\mu) 
\,. \label{eq:exp2} 
\end{eqnarray}
One may ask whether the first $f$-derivative 
of the orbital positional element $d\phi_C/df$ may become large 
for large $\tilde t(f)$, invalidating the expansion. 
However, as we will see in the next section, 
this is not a problem.

In the adiabatic expansion we regard the scalar field as a function of
these orbital elements and we expand the field as
\begin{eqnarray}\label{eq:adiabexp}
\Phi\left(x | C(f(x^\alpha))\right) 
= \Phi^{(0)}\left(x^\alpha | C(f(x^\alpha))\right) 
+\mu \Phi^{(1)}\left(x^\alpha | C(f(x^\alpha))\right)
+\mu^2 \Phi^{(2)}\left(x^\alpha | C(f(x^\alpha))\right) +\cdots \,. 
\end{eqnarray}
To see how this expansion is used, 
we consider the wave operator acting on the scalar field. 
When the differential operator acts 
on the orbital elements of the scalar field, 
it results in a factor $\mu$ 
according to the scheme of (\ref{eq:exp}). 
To the leading order (i.e., order $\mu^0$) of the adiabatic expansion, 
we have the equation for $\Phi^{(0)}$ to be 
\begin{eqnarray} 
\left[\Box \Phi^{(0)} (x^\alpha| C)\right]_{C=C(f(x^\alpha))} 
= \rho \left(x^\alpha | C(f(x^\alpha))\right) 
\,. \label{eq:adi_e} 
\end{eqnarray} 
On the left hand side of (\ref{eq:adi_e}), 
the wave operator 
is considered to act 
only on the direct dependence on $x^\alpha$, not on the dependence implicit
in the dependence of $C$ on $f$. In this sense, then, the 
orbital elements can be considered to be constants.
By also treating the $C(f)$ on the right as if they were constants, we
interpret the wave equation (\ref{eq:adi_e}) as 
\begin{eqnarray} 
\Box \Phi^{(0)} (x | C) = \rho \left(x | C)\right) 
\,, \label{eq:adi_ex} 
\end{eqnarray} 
for which we can give 
the well known solution for a circular orbit 
with constant orbital elements, 
\begin{eqnarray}
\Phi^{(0)} \left(x^\alpha| C(f(x^\alpha))\right) 
&=& q \sqrt{1-r_C^2 \Omega^2} \sum_{lm} (-im\Omega)
\biggl(h_l^{(1)}(m\Omega r)j_l(m\Omega r_C) \theta(r-r_C)
\nonumber \\ && \qquad\qquad\qquad 
+j_l(m\Omega r)h_l^{(1)}(m\Omega r_C) \theta(r_C-r)\biggr)
Y_{lm}(\theta,\phi) Y^*_{lm}(\pi/2,0) e^{-im(\Omega t +\phi_C)}
\,, \label{eq:adi_f}
\end{eqnarray}
where the orbital elements are evaluated 
by $r_C=r_C(f(x^\alpha))$, $\Omega=\Omega(f(x^\alpha)))$ 
and $\phi_C=\phi_C(f(x^\alpha)))$. 

We note that, to this leading order of the adiabatic expansion, 
the foliation function does not play a crucial role 
in the practical calculation, 
but it does play an important role 
in the meaning of the expansion, 
in particular how the evolving elements of the orbit 
affect the solution at points off the orbit. 
It tells us that the solution at field point $x^\alpha$ 
is the circular orbit solution 
for the values of $r_C$, $\Omega$ and $\phi_C$ 
that occur at the ``time'' $f(x^\alpha)$ on the orbit.

\section{First Post-Adiabatic Field}
\label{sec:adi1}

In the field equation (\ref{eq:adi_ex}), 
we ignored the derivative acting on the orbital elements 
and, by the rule (\ref{eq:exp}), 
we found a field $\Phi^{(0)}$ that satisfies (\ref{eq:feq}) 
only to accuracy  $O(\mu)$. 
We now turn to the calculation of the solution to the next 
order in $\mu$.

Using the orders for the derivatives of the orbital elements 
(\ref{eq:exp1}) and (\ref{eq:exp2}) 
we can substitute the adiabatic expansion (\ref{eq:adiabexp}) 
into the field equation (\ref{eq:feq}), 
and we can extract the terms of order $\mu$. 
The details are given in Appendix A, 
and the result is a field equation 
for the first-order post-adiabatic field $\Phi^{(1)}$: 
\begin{eqnarray}
\left[\Box \Phi^{(1)} (x|C) \right]_{C=C(f)} 
&=& \rho^{(1)} \left(x|C(f)\right) 
\,, \label{eq:1adi_e} \\ 
\rho^{(1)}\left(x|C(f)\right) 
&=& -\sum_{C^a = r_C,\Omega,\phi_C}\biggl\{
2\left[\partial_\alpha \partial_a \Phi^{(0)}(x|C)\right]_{C=C(f)}
g^{\alpha\beta} \partial_\beta f 
+\partial_a \Phi^{(0)}\left(x|C(f)\right)
\Box f \biggr\}\left[{dC^a \over df}\right]^{(1)}
\nonumber \\ && 
-\partial_{\phi_C} \Phi^{(0)}\left(x|C(f)\right)
g^{\alpha\beta}  \partial_\alpha f \partial_\beta f 
\left[{d^2\phi_C \over df^2}\right]^{(1)}
\,, \label{eq:1adi_s} 
\end{eqnarray} 
where the notation $[dC/df]^{(1)}$ and $[d^2C/df^2]^{(1)}$, 
means that these derivatives are 
only to be taken to first order in $\mu$.
On the left hand side of (\ref{eq:1adi_s}) 
the derivatives in the $\Box$ operator are understood 
only to act on the explicit $x^\alpha$ dependence, 
and not on the orbital elements. 
The first-order nature of the left hand side is due to the fact 
that $\Phi^{(1)}$ itself is defined 
to be first order in (\ref{eq:adiabexp}).  
The result in (\ref{eq:1adi_s}), with the orbital elements 
treated as effective constants on the left 
is the scalar version of the post-adiabatic equation 
first derived in Ref.~\cite{adi}. 

The foliation function $f(x^\alpha)$ played only a passive role in the
adiabatic approximation of Eqs.~(\ref{eq:adi_e})-- (\ref{eq:adi_f}).
By contrast, here the post-adiabatic terms depend explicitly on the
choice of $f(x^\alpha)$.  The choice of the foliation function is
similar to a gauge choice, in that the value of any physical quantity
cannot depend on the choice\cite{eannaprivate}.  As a useful check,
the foliation-invariance of the first post-adiabatic approximation
field $\Phi^{(1)}$ is verified in Appendix \ref{app:inv}.

Though the choice of foliation function is arbitrary in principle, an
inappropriate choice will make the first post-adiabatic term
$\Phi^{(1)}$ divergent so that a regularization procedure will be
needed to extract physical information.  In Appendix \ref{app:1asym},
we study the asymptotic behavior of the source for the first
post-adiabatic term and we find that, if the foliation is
asymptotically null, the first post-adiabatic term behaves as
$\Phi^{(1)} \to O(1/r)$ at large radius.

Another insight on the appropriateness of a null foliation comes from
the retarded time Lienard-Wiechert solution \cite{mathewswalker} to
(\ref{eq:feq}):
\begin{equation}\label{eq:LWexact}
\Phi=-\,\frac{q}{4\pi}\,\frac{1}
{
|\vec{x}-\vec{\xi}(t')|-\dot{\vec{\xi}}({t'})
  \cdot(\vec{x}-\vec{\xi}(t'))   
}
\end{equation}
where ${\vec\xi({\tau}})$ has the components $\tilde{z}^\alpha(\tau)$
of the orbit of the source particle, and where $t'$ is the retarded 
time, a solution of
\begin{equation}\label{eq:retard}
t'=t-|\vec{x}-  \vec{\xi}(t')  |\ .
\end{equation}
For
the evolving equatorial orbital motion
described by  parameters $\Omega(\tau)$, $r_C(\tau)$, and
$\phi_C(\tau)$, eq.~(\ref{eq:retard}) becmes
\begin{equation}\label{eq:retard1}
t'=t-\sqrt{
r^2+r_C^2-2rr_C\sin\theta\cos{(\Omega(t') t'+\phi_c(t')-\phi)}
\;}
\ .
\end{equation}
The exact solution of (\ref{eq:retard1}) for $t'$ as a function of
$t,r,\theta,\phi$ is a null foliation, and would, in a sense, be the
perfect foliation for the problem since with this foliation the
solution (i.e.\,, that of (\ref{eq:LWexact})\,) is guaranteed to have
no apparent singularities except at the source points.
Equation (\ref{eq:retard1}) cannot in general  be solved 
exactly, but for $r\gg r_C$, and $|t-r|\gg r_C$, 
the approximate solution is a simpler null foliation
$t'=t-r$.

For simplicity, as well as from the form of the Lienard-Wiechert solution,
we choose the foliation
\begin{eqnarray}
f(x^\alpha) = t -r \,. \label{eq:fol} 
\end{eqnarray}
With this choice, $\Box f=-2/r$ in the second term of
(\ref{eq:1adi_s}),  $g^{\alpha\beta}
\partial_\alpha f\partial_\beta f=0$
in the last term, 
the time
coordinate of the orbit is given as $\tilde t(f) = f +r_C(f)$,
and the evolution of $\phi_C$ is given by
\begin{eqnarray}
{d \over df} \phi_C 
= -\left[t\!-\!r+r_C(f)-r_C(f\!=\!0)\,\right]\, {d \over df}\Omega 
\,. \label{eq:eq:phiff} 
\end{eqnarray}
This puts the source term of (\ref{eq:1adi_s}) in the form 
\begin{eqnarray}
\rho^{(1)} \left(x | r_C,\Omega,\phi_C\right) 
&=& 2 \left[ \left(\partial_t+{1 \over r}\partial_r r\right)
{\partial \over \partial C^a}\Phi^{(0)}(x|C) \right]_{C=C(f)}
\left[{d C^a \over df}\right]^{(1)} 
\nonumber \\ 
&=& q \sqrt{1-v^2} \sum_{lm} (-im\Omega)^2 
\Biggl\{S^{(+)}_{lm}\theta(r-r_C)
+S^{(-)}_{lm}\theta(r_C-r)+S^{(p)}_{lm}r_C\delta(r-r_C)\Biggr\}
\nonumber \\ && \qquad \qquad \qquad 
\times Y_{lm}(\theta,\phi) Y^*_{lm}(\pi/2,0)
e^{-im(\Omega t +\phi_C)}
\left[{dv \over df}\right]^{(1)}\label{eq:rh01}
\,, 
\end{eqnarray}
where 
\begin{eqnarray}
S^{(+)}_{lm} &=& S^{(1)j}_{lm}\left[h^{(1)}_l(z)
+{i \over z}{d \over dz}zh^{(1)}_l(z)\right]_{z=m\Omega r}
\nonumber \\ && 
+S^{(2)j}_{lm}\left[2h^{(1)}_l(z)+2z{d \over dz}h^{(1)}_l(z)
-i2zh^{(1)}_l(z)+i{l(l+1) \over z}h^{(1)}_l(z)\right]_{z=m\Omega r}
\,, \\ 
S^{(-)}_{lm} &=& S^{(1)h}_{lm}\left[j_l(z)
+{i \over z}{d \over dz}zj_l(z)\right]_{z=m\Omega r}
\nonumber \\ && 
+S^{(2)h}_{lm}\left[2j_l(z)+2z{d \over dz}j_l(z)
-i2zj_l(z)+i{l(l+1) \over z}j_l(z)\right]_{z=m\Omega r}
\,, \\ 
S^{(p)}_{lm} &=& -{2\over m^2 v^3} 
\,, \\ 
S^{(1)k}_{lm} &=& \left\{ -{v \over 1-v^2}+{3 \over v}
+3im\left(1-{v^2\over v^2(f=0)}\right) \right\} k_l(mv)
+m \left({d \over dz}k_l(z)\right)_{z=mv} 
\,, \\ 
S^{(2)k}_{lm} &=& {3 \over v} k_l(mv)
\,,\label{eq:Sterms}
\end{eqnarray}
with $k$ representing either $h^{(1)}$ or $j$. 

For the most part, the derivation of
(\ref{eq:rh01}) -- (\ref{eq:Sterms}) is straightforward, but there is
at least one point worth mentioning: The derivative with respect to
$\phi_C$ entails a factor, shown in (\ref{eq:eq:phiff}), of 
$
t-r+r_C(f)-r_C(f\!=\!0)
$.
The leading $t$ will cancel with the $t$ arising from the derivative 
of $e^{-im\Omega(f)t}$, with respect to $f$, and the $-r$ is rewritten
as $-m\Omega z$, leaving us with 
\begin{equation}
r_C(f)-r_C(f\!=\!0)=\Omega^{-1}v\left[1-\frac{r_C(f\!=\!0)}{r_C(f)}\right]
=\Omega^{-1}v\left[1-\frac{v^2(f)}{v^2(f\!=\!0)}\right]\,.
\end{equation}
Here the relationship $r_C=M/v^2$, from (\ref{eq:orbx}), has been 
used and, as usual, the symbols $\Omega$ and $v$, with no explicit argument
are understood to be $\Omega(f)$ and $v(f)$.

For the source in (\ref{eq:rh01}) -- (\ref{eq:Sterms}), the solution
of (\ref{eq:1adi_e}) for $\Phi^{(1)}$ can be found using the standard 
retarded Green function for the wave operator. This solution 
is summarized in Appendix
\ref{app:int}.  It should be noted that the source term above has a
discrete spectrum and contains only Fourier components with
$\omega=m\Omega$ and that each $(lm)$-mode of $S^k_{lm}$ contributes
only to the $(lm)$-mode of $\Phi^{(0)}$.  This is due to our
spherically symmetric choice of the foliation in (\ref{eq:fol}).  With
this choice, no explicit angular dependence is introduced into the
differential operators.  The separation of Fourier modes stops when we
come to the explicit evaluation of $[dv/df]^{(1)}$.  This is a
nonlinear step that mixes the contributions from all Fourier modes.

\section{Summary and discussion}
\label{sec:sum}

We have applied the two-timescale method to the simple well-understood
problem of a slowly evolving quasi-circular orbit of a scalar charged
particle in a central potential. The question motivating this work 
was that of the justification, and applicability of the adiabtic 
approximation and its extension. We have focused on the
manner in which the slowly changing orbital elements of the
trajectory, the radius $r_C$, angular velocity $\Omega$, and phase
constant $\phi_C$ are to be promoted to spacetime fields through the
choice of some foliation function $f(x^\alpha)$. We have shown that
this choice, which plays only a passive role in the lowest-order adiabatic
expansion, is crucial for the post-adiabatic
fields.  
In particular, we show that 
the first-order adiabatic field
$\Phi^{(1)}$ is well behaved as $r\rightarrow\infty$ if $f$
is chosen to be $t-r$, but 
 is not well behaved
for a general choice of foliation function, in particular for the 
acausal choice $f=t$, which has the orbital elements change 
throughout spacetime according to coordinate time. 

An advantage of our simple model is that we have an alternative approach 
to the description of the field, the Lienard-Wiechert solution.
We can get insight into the pathology of the wrong foliation by 
considering the $r\gg r_C$ form of the Lienard-Wiechert solution 
in (\ref{eq:LWexact})
\begin{equation}\label{eq:LWfar}
\Phi=-\,\frac{q}{4\pi r}\,\frac{1}
{1-
\dot{\vec\xi}({t'})
\cdot\vec{x}/r   
\ }\,.
\end{equation}
We have already argued, on the basis of the retardation equation
(\ref{eq:retard}) or (\ref{eq:retard1}), that $f=t-r$ is the correct
large $r$, foliation function, but now we can consider what would
happen for the foliation choice $t'=t$. In this case, at large $r$,
the value of $\dot{\vec\xi}({t'=t})$ would differ from the correct
value $\dot{\vec\xi}({t'=t-r})$, by an amount that is proportional to
$r$ and thus would constitute a contribution to the field that does
not fall off as $1/r$.

It should be noted that the Lienard-Wiechert solution is itself a
demonstration that for our simple problem the adiabatic approximation
can be made to be exact over an arbitrarily long timescale.  In
(\ref{eq:LWexact}) one needs only to take the exact solution for $t'$
as the foliation and to describe the motion $\vec{\xi}(\tau)$ in terms
of evolving orbital constants, a description that can always be given
if enough constants are included.  From this point of view, adiabatic
expansions of different orders represent various accuracies with which
the foliation function $f$ is equal to the $t'$ that solves
(\ref{eq:retard}).  In our equatorial orbit problem, the usual lowest
order adiabatic approximation (say that of (\ref{eq:adi_f}))
corresponds to the following procedure: One fixes the orbital
constants $\Omega$, $r_C$ and $\phi_C$, then solves symbolically for
$t'(x^\alpha)$ parameterized by these constants and writes the
Lienard-Wiechert solution in terms of that solution for $t'$. In this
solution, one then promotes the ``constants'' to functions of whatever
foliation is chosen.

The Lienard-Wiechert solution, of course, applies only to problems
involving linear fields obeying the flat spacetime wave equation.
The two-timescale approach, in principle, applies much more generally.
But at least 
four important questions remain about the two-timescale
approach.  The first is whether it converges for our simple problem
(or similar linear problems). We feel that the existence and form of the
Lienard-Wiechert solution all but guarantees that the post-adiabatic
expansion, {\em if done correctly}, must converge.  The caveat here,
that the expansion must be done correctly, points to the need for the
correct choice of foliation. A general argument based on
(\ref{eq:LWfar}), and the details of the post-adiabatic equation in
(\ref{eq:2adi_s}) and its higher order extensions, both point to the
choice $f=t-r$ being adequate to all orders.

The second important question about the expansion is whether it
converges for fields in curved spacetimes, such as black hole
spacetimes.  The issue is the choice of the foliation around the black
hole horizon and the convergence of the post-adiabatic expansion.  The
curved-space generalization of the Lienard-Wiechert solution is
obtained by the formal Hadamard series expansion using the bi-tensor
formalism\cite{Hadamard}.  On the basis of this we suspect that the
post-adiabatic expansion must converge with the choice of null
foliation to the future horizon, but this is a question that deserves
further study.

The third important question about the expansion is the issue of the
gauge condition when we apply this expansion to the gravitational
perturbation problem\cite{adi}.  In Sec.~\ref{sec:adi0}, we found that
the first $f$-derivative of the positional element (\ref{eq:exp2})
could be large for large $\tilde t(f)$.  However, as we see from
(\ref{eq:rh01}), this does not invalidate the post-adiabatic expansion
in our simple scalar model.  This is not the case for gravitational
perturbations and it is necessary to choose an appropriate gauge
condition for the post-adiabatic expansion so that gravitational
perturbations become valid on a longer time scale\cite{adi}.

The last important question about the expansion is whether it
converges for more realistic problems in relativity, especially
nonlinear problems including the dynamics of geometry.  Here we feel
that the issue of divergences at large-$r$ will probably be similar or
identical to that for the linear problems, since the fields are weak
at large $r$.  The issue of the post-adiabatic expansion near strong
field sources is one that will have to await the exploration of a
nonlinear model.  A point deserving particular attention is that the
dynamical horizon may complicate the analysis and the convergence due to the time
evolution of the black hole mass.  To get some insights about
nonlinear issues, we hope to apply the two-timescale method to is a
post-helical solution based on a helical nonlinear problem, such as
that in the post-Minkowski solution of Beetle {\it et
al.}\cite{pMpaper}

\section{Acknowledgment}
We gratefully acknowledge support for this work under NSF grants
 PHY-0601459 and PHY-0554367, by the Brinson Foundation, and by the
 Center for Gravitational Wave Astronomy. We thank \'Eanna Flanagan
 for helpful suggestions.

\appendix

\section{Adiabatic Expansion of the wave operator}
\label{app:exp} 

In this appendix, we consider the two-timescale expansion of the wave
operator.  For the application of the two-timescale expansion, we regard
the scalar field $\Phi$ as a function of the spacetime coordinates
$\{x^\alpha\}$ and of the orbital elements, $\Omega,r_C$ and $\phi_C$,
collectively denoted $C^a$. In addition to its direct dependence on the 
spacetime coordinates, the scalar field has an additional dependence
through the foliation function $f(x^\alpha)$:
\begin{eqnarray}
\Phi &=& \Phi \left( x | C(f) \right) \,. 
\end{eqnarray}
As explained in (\ref{eq:exp}), in the two-timescale expansion we assume
that the derivative of the orbital principal elements is small, 
and in particular that
\begin{eqnarray}
{d^n \over df^n}v \sim O(\mu^n) \,, \label{eq:ansatz} 
\end{eqnarray}
where $\mu$ is the ratio of dynamic to radiation-reaction timescales.
The derivative of the orbital positional elements is 
obtained from the orbita equation of motion. 

It is crucial to deal separately
with the coordinate derivative of the scalar field with fixed $C^a$ 
and with the coordinate derivative of the scalar field 
due to the spacetime dependence implicit in $C^a(f(x^\alpha))$.
For  clarity of the notation, 
we use 
\begin{eqnarray}
\left[\partial_\alpha \Phi (x|C) \right]_{C=C(f)} \,, 
\end{eqnarray}
to denote the derivative with respect to the cordinates 
with the orbital elements $C^a$ fixed. 
We also use 
\begin{eqnarray}
\partial_a \Phi 
:= \left[{\partial \over \partial C^a} \Phi (x|C) \right]_{C=C(f)} 
\end{eqnarray}
to denote the derivative 
with respect to the orbital elements $C^a$. 

With these notations, 
the partial derivative of the scalar field 
is written as 
\begin{eqnarray}
\partial_\alpha \Phi \left(x | C(f)\right) 
&=& \left[\partial_\alpha \Phi\right]_{C=C(f)} 
+ \partial_a \Phi {dC^a \over df} \partial_\alpha f \,. 
\end{eqnarray}
The second partial derivative becomes 
\begin{eqnarray}
\partial_\alpha \partial_\beta \Phi \left(x | C(f)\right) 
&=& \left[\partial_\alpha \partial_\beta \Phi \right]_{C=C(f)} 
+ \left[\partial_a \partial_\alpha \Phi \right]_{C=C(f)} 
{dC^a \over df} \partial_\beta f 
+ \left[\partial_a \partial_\beta \Phi\right]_{C=C(f)} 
{dC^a \over df} \partial_\alpha f 
\nonumber \\ && \qquad
+ \partial_a \partial_b \Phi 
{dC^a \over df}{dC^b \over df} \partial_\alpha f \partial_\beta f
+ \partial_a \Phi \left(
{dC^a \over df} \partial_\alpha \partial_\beta f
+{d^2 C^a \over df^2} \partial_\alpha f \partial_\beta f
\right) \,, 
\end{eqnarray}
where we use the fact that 
the coordinate derivative with the fixed orbital elements 
commute with the derivative with respect to the orbital elements , i.e.\,, 
$\left[\partial_a \partial_\alpha \Phi\right]_{C=C(f)} 
= \left[\partial_\alpha \partial_a \Phi\right]_{C=C(f)}$. 

Using the results above, we have 
that in general spacetime coordinates
the wave operator acting on the scalar field is
\begin{eqnarray}
\Box\Phi &=& \left[\Box\Phi\right]_{C=C(f)}
\nonumber \\ && 
+ g^{\alpha\beta}\left(
2\left[\partial_\alpha \partial_a \Phi\right]_{C=C(f)}
\partial_\beta f
+\partial_a \Phi \partial_\alpha \partial_\beta f
-\Gamma^\gamma_{\alpha\beta}
\partial_a \Phi\partial_\gamma f \right){dC^a \over df}
\nonumber \\ && 
+ \left(\partial_a \partial_b \Phi{dC^a \over df}{dC^b \over df}
+ \partial_a \Phi {d^2C^a \over df^2}\right)
g^{\alpha\beta}\partial_\alpha f \partial_\beta f 
\,, \label{eq:wave}
\end{eqnarray}
where $\Gamma^\gamma_{\alpha\beta}$ is the Christoffel symbol. 

In applying (\ref{eq:wave}) to find adiabatic fields of any order, one
must count the order by adding the index $n$ of the field $\Phi^{(n)}$
to the number of $C^a$ factors, keeping in mind that $d^2C^a/df^2$ can
be first or second order. Thus, for example,
$g^{\alpha\beta}\partial_\alpha\partial_a\Phi^{(2)}(dC^a/df)f_{,\beta}$ 
is third order 
and contributes only to the source for $\Box\Phi^{(3)}$, while 
$\partial_\alpha\partial_a\Phi^{(2)}(d^2C^a/df^2)
g^{\alpha\beta}f_{,\alpha}f_{,\beta}$ is a source for both
$\Box\Phi^{(3)}$ and 
$\Box\Phi^{(4)}$.

To first order, the source that result from 
(\ref{eq:wave}) is that in
(\ref{eq:1adi_s}). To second order, the result is
\begin{eqnarray}
\left[\Box \Phi^{(2)} (x|C) \right]_{C=C(f)} 
&=& \rho^{(2)} \left(x|C(f)\right) 
\,, \label{eq:2adi_e} \\ 
\rho^{(2)}\left(x|C(f)\right) 
&=& -\left(\partial_a \partial_b \Phi^{(0)}
\left[{dC^a \over df}\right]^{(1)}\left[{dC^b \over df}\right]^{(1)}
+\partial_a \Phi^{(0)}\left[{d^2C^a \over df^2}\right]^{(2)}
\right)g^{\alpha\beta}\partial_\alpha f\partial_\beta f
\nonumber \\ && 
-\biggl\{g^{\alpha\beta} 
2\left[\partial_\alpha \partial_a \Phi^{(1)}(x|C)\right]_{C=C(f)}
\partial_\beta f 
+\partial_a \Phi^{(1)}\left(x|C(f)\right)\Box f
\biggr\}
\left[{dC^a \over df}\right]^{(1)}
\nonumber\\&&
-\partial_a\Phi^{(1)} \left[{d^2C^a \over df^2}\right]^{(1)}
g^{\alpha\beta}\partial_\alpha f\partial_\beta f \,.\label{eq:2adi_s}
\end{eqnarray} 
It should be noted that for null foliations the factor $g^{\alpha\beta}
f_\alpha f_\beta$ vanishes, simplifying the results.

\section{Invariance of the first post-adiabatic expansion}
\label{app:inv}

Our adiabatic expansion uses a foliation function that is arbitrary
in principle.  If the result of the adiabatic expansion depends on the
choice of this foliation, it means there must be a physical condition
to determine the foliation.  Otherwise, the adiabatic expansion would
not have physical predictability.  In this appendix, we shall prove
that the field found by the adiabatic expansion does not depend on the
choice of the foliation to the first post-adiabatic expansion, and
therefore, at least to this order, does have physical predictablity.

We consider an infinitestimal change in the foliation function 
\begin{eqnarray}\label{eq:deltaf}
f(x^\alpha) \rightarrow f(x^\alpha)+\delta f(x^\alpha) \,, 
\end{eqnarray}
and we show here that the first
post-adiabatic field $\Phi^{(1)}$
does not change to first order in $\delta f$.

There are two constraints on our choice of $f$, and hence on $\delta
f$.  First, we recall the role of the foliation function. The orbital
elements $C$, that are defined by the evolving orbit are promoted to
spacetime functions by making them functions $C(f(x^\alpha))$. We
choose to leave consant the value of $f$ on every point of the
orbit. (To do otherwise would not really involve a change in the
foliation, but rather would be a change in the parameterization of the
orbit that would change the functional forms of the $C(f)$).  This
gives us the constraint
\begin{eqnarray}
\delta f(x^\alpha) = 0 \,,\quad \mbox{for $x^\alpha$ on the orbit}
\,. \label{eq:const1} 
\end{eqnarray}
Our second constraint is that we choose $\delta f$ to behave 
at infinity so that
\begin{eqnarray}
{\partial\,\delta\! f\over \partial x^\alpha} \rightarrow \mbox{constant} 
\,,\quad \mbox{as } r \rightarrow \infty 
\,. \label{eq:const2} 
\end{eqnarray}
The reason for this, to be clarified below, is connected to the large-$r$ boundary
conditions that determine the solution for 
$\Phi^{(1)}$.

The differential change $\delta f$ affects the field
$\Phi=\Phi^{(0)}+\Phi^{(1)} +\cdots$ in two ways. First, at a given
spacetime point $x^\alpha$ the field formally denoted as $\Phi=\Phi^{(0)}$
will have a change of order $\mu$ according to 
\begin{equation}
\Phi^{(0)}\left(x|C(f)\right) \rightarrow
\Phi^{(0)}\left(x|C(f+\delta f)\right) 
= \Phi^{(0)}\left(x|C(f)\right) 
+ \delta \Phi^{(0)}\left(x|C(f)\right) 
\,, 
\end{equation}
where
\begin{equation}
\delta \Phi^{(0)}\left(x|C(f)\right) 
=\partial_a \Phi^{(0)} {dC^a \over df} \delta f
\,. \label{eq:d_0adi} 
\end{equation}
Because $dC^a/df$ is $O(\mu)$ 
by the ansatz of the two-timescale expansion (\ref{eq:ansatz}), 
$\delta \Phi^{(0)}$ is regarded 
as the part of the first post-adiabatic term. 
We note that 
\begin{equation}
\left[\Box\delta\Phi^{(0)}\right]_{C\!=\!C(f)}=\Box\left[
\partial_a \Phi^{(0)} {dC^a \over df} \delta f\right]
=\left[2g^{\alpha\beta}\partial_a\partial_\alpha\Phi^{(0)}\delta_\beta f
+\partial_a\Phi^{(0)}
\Box\delta f\right]{dC^a \over df}\label{boxdeltaPhi1}\,.
\end{equation}
The term $\partial_a\Box\Phi^{(0)}\delta f (dC^a/df)$
has been omitted, since $\Box\Phi^{(0)}$ vanishes except on the orbit, where,
by constraint (\ref{eq:const1}) the factor $\delta f$ vanishes.

To find the change in the 
formal post-adiabatic term
$\Phi^{(1)}$ we consider
 the transformation of $\rho^{(1)}$ 
\begin{eqnarray}
\rho^{(1)} &\rightarrow& \rho^{(1)}+\delta \rho^{(1)}
\,. 
\end{eqnarray}
We are interested in $\delta\rho^{(1)}$, 
the change, to first order in $\mu$, to the right hand side of 
(\ref{eq:rh01}) induced by $\delta f$.
It is crucial to see that 
$\delta \rho^{(1)}$ does not come from changes
in $\Phi^{(0)}$ nor to $[dC^a/df]^{(0)}$
in that right hand side; such changes would be  
post-adiabatic terms higher order in $\mu$.
To first order in $\mu$ the only changes induced by 
the change (\ref{eq:deltaf}) are those in
\begin{eqnarray}
\delta \rho^{(1)}&=& -\biggl\{
2\left[\partial_\alpha \partial_a \Phi^{(0)}(x|C)\right]_{C=C(f)}
g^{\alpha\beta} \partial_\beta \delta f 
\nonumber \\ && \qquad\quad 
+\partial_a \Phi^{(0)}\left(x|C(f)\right)
\Box \delta f \biggr\}\left[{dC^a \over df}\right]^{(0)} 
\,. \label{eq:d_1adi_s} 
\end{eqnarray}
 
In principle, we can find $\delta\Phi^{(1)}$
from
\begin{eqnarray}
\left[\Box \delta \Phi^{(1)} \right]_{C\!=\!C(f)} 
&=& \delta \rho^{(1)} 
\,. 
\end{eqnarray}
Since the right hand sides of (\ref{eq:d_1adi_s}) and
(\ref{boxdeltaPhi1}) add to zero, we have 
\begin{eqnarray}
\left[\Box \left(\delta\Phi^{(1)}+\delta\Phi^{(0)}\right) \right]_{C\!=\!C(f)} 
&=&0\label{eq:waveqdeltaPhi}
\,,
\end{eqnarray}
so the wave equation is sourceless for the change $\delta\Phi$ of the
scalar field, to first adiabatic order. 

In order to conclude that the change $\delta\Phi$ is zero to first
order, we need to rule out ``free wave'' solutions of
(\ref{eq:waveqdeltaPhi}).  This can be accomplished by imposing the
constraint in (\ref{eq:const2}).  If some such constraint is not
imposed then nonzero solutions for $\delta\Phi$ exist, and a change in
the foliation function appears to make a change, to first adiabatic
order, in the solution for $\Phi$.  But the change in the foliation
function would entail a change from a solution that is causal, i.e.,
one that has only outgoing waves, to one that is not causal. In the
adiabatic expansion we use only causal Green functions to solve for
the various orders of $\Phi^{(n)}$.

\section{Asymptotic behavior of the source term 
for the first post-adiabatic term}
\label{app:1asym}

In this appendix, we consider a general spherical foliation 
defined as 
\begin{eqnarray}
f =f(t,r) \,, 
\end{eqnarray}
and we investigate the asymptotic behavior of the source term 
for the first post-adiabatic field
with this general foliation function. 
We use the asymtotic form of the adiabatic scalar field 
written as 
\begin{eqnarray}
\Phi^{(0)}\left(x|C(f)\right) &=& {1 \over r}\sum_{lm;n=0,1,2,\cdot}
A^{(n)}_{lm}\left(C(f)\right)\,Y_{\ell m}(\theta,\phi)
{1 \over r^n}e^{-im\left(\Omega(f)(t-r)+\phi_C(f)\right)} \,. 
\end{eqnarray}
(See (\ref{eq:adi_f}) for the full expression.)

The source term for the first post-adiabatic term, for the spherically
symmetric foliation is
\begin{eqnarray}
\rho^{(1)}(x;C) &=& \sum_{C^a=\Omega,\phi_C}\Biggl\{
\left[\left(2\partial_t f \partial_t -2\partial_r f \partial_r 
-{2 \over r}\partial_r f +\partial^2_t f-\partial^2_rf\right) 
{\partial \over \partial C^a}\Phi^{(0)}(x,C)\right]
\left[{d C^a \over df}\right]^{(1)}
\nonumber \\ 
&& \qquad \qquad \quad 
+\left((\partial_t f)^2-(\partial_t f)^2\right)
{\partial \over \partial C^a}\Phi^{(0)}(x,C)
\left[{d^2 C^a \over df^2}\right]^{(1)}\Biggr\}\label{eq:Brho1}
\,, 
\end{eqnarray}

The regular behavior of $\Phi$, and hence of the first post-adiabatic
term $\Phi^{(1)}$ is $\sim O(r^{-1})$ at large $r$.  For this to be
the case, the first-order source term $\rho^{(1)}$ must fall off at
large $r$ as $O\left(r^{-3}\right)$\,,
and this condition constrains the foliation function.
To order $1/r$, in the expression above
\begin{equation}
2\partial_t f \partial_t-2\partial_r f \partial_r
=-2im(\partial_tf+\partial_rf)\,.
\end{equation}
The function 
$\partial_a C^a\Phi^{(0)}(x,C)$ 
falls off as $1/r$, hence $\rho^{(1)}$ is asymptotically $1/r$
unless the leading terms vanish, that is unless
\begin{eqnarray}
0 = \partial_t f +\partial_r f = \partial_t^2 f -\partial_r^2 f 
\,, 
\end{eqnarray}
which indicates that the foliation surface 
must asymptotically be a future null cone. 
By these conditions, we have the foliation function written as 
\begin{eqnarray}
f(t,r) \to f(t-r) 
\quad \hbox{at} \quad r \to \infty \,. 
\end{eqnarray}
If we put $f=t-r$ in (\ref{eq:Brho1}) we get
\begin{equation}
\rho^{(1)}(x;C) = 2\sum_{C^a=\Omega,\phi_C}
\left[\left(\partial_t + \partial_r 
+{1 \over r} \right) 
{\partial \over \partial C^a}\Phi^{(0)}(x,C)\right]
\left[{d C^a \over df}\right]^{(1)}\,.
\end{equation}
Since
\begin{equation}
\partial_a C^a\Phi^{(0)}(x,C)=
\frac{1}{r}e^{-im\left(\Omega(f)(t-r)+\phi_C(f)\right)}
\bigl[{\cal A}(C^a,\theta,\phi)+\frac{B{(C^a,\theta,\phi)}}{r}+\cdots\biggr]\,,
\end{equation}
we have that $\rho^{(1)}(x;C)$ falls off as $r^{-3}$, as it must.

\section{Integral formula for the first post-adiabatic term}
\label{app:int}

We summarize the integral formula for the field equation 
under the outgoing-wave boundary condition. 
We consider the scalar wave equation 
\begin{eqnarray}
\Box  F(x^\alpha) &=& S(x^\alpha) \,, 
\end{eqnarray}
for flat spacetime,
i.e., $\Box = -\partial_t^2 +\partial_x^2 +\partial_y^2 +\partial_z^2$. 
In spherical coordinates, 
it is covenient to use the  decomposition 
\begin{eqnarray}
F(x) = \sum_{lm\omega}F_{lm\omega}(r)
Y_{lm}(\theta,\phi)e^{-i\omega t} \,, \quad 
S(x) = \sum_{lm\omega}S_{lm\omega}(r)
Y_{lm}(\theta,\phi)e^{-i\omega t} \,.
\end{eqnarray}
A formal solution of the radial mode functions 
is obtained by the Green's method as 
\begin{eqnarray}
F_{lm\omega}(r) &=& \int dr' g_{lm\omega}(r,r') S_{lm\omega}(r) 
\,, \\
g_{lm\omega}(r,r') &=& -i \omega r'^2 
\left(h^{(1)}_l(\omega r)j_l(\omega r')\theta(r-r')
+j_l(\omega r)h^{(1)}_l(\omega r')\theta(r'-r)\right) \,, 
\end{eqnarray}
where $\theta(x)$ is the step function, 
i.e. $\theta(x)=1$ when $x>0$ and $\theta(x)=0$ otherwise. 

For the calculation of the first post-adiabatic term, 
we write the source term schematically as 
\begin{eqnarray}
S_{lm\omega}(r) &=& s^p_{lm\omega}r_C\delta(r-r_C)
\nonumber \\ && 
+\Biggl\{
s^{h(1)}_{lm\omega}\left[h^{(1)}_l(z)
+{1 \over z}{d \over dz}zh^{(1)}{}'_l(z)\right]_{z=\omega r}
+s^{h(2)}_{lm\omega}\left[h^{(1)}_l(z)
+z {d\over dz}h^{(1)}_l(z)\right]_{z=\omega r}
\nonumber \\ && \qquad \qquad 
+s^{h(3)}_{lm\omega}\left[-i2z h^{(1)}_l(z)
+i{l(l+1) \over z}h^{(1)}_l(z)\right]_{z=\omega r}
\Biggr\}\theta(r-r_C) 
\nonumber \\ && 
+\Biggl\{
s^{j(1)}_{lm\omega}\left[j_l(z)
+{1 \over z}{d \over dz}zj_l(z)\right]_{z=\omega r}
+s^{j(2)}_{lm\omega}\left[j_l(z)
+z {d\over dz}j_l(z)\right]_{z=\omega r}
\nonumber \\ && \qquad \qquad 
+s^{j(3)}_{lm\omega}\left[-i2z j_l(z)
+i{l(l+1) \over z}j_l(z)\right]_{z=\omega r}
\Biggr\}\theta(r_C-r) 
\,, \label{eq:sxx}   
\end{eqnarray}
where $s^{h/j(1/2/3)}_{lm\omega}$ does not depend on $r$. 

The radial mode function 
induced by the first term on the right hand side of (\ref{eq:sxx}) 
is obtained in a trivial manner as 
\begin{eqnarray}
F_{lm\omega}^p(r) 
&=& \int dr' g_{lm\omega}(r,r') 
s^p_{lm\omega}r_C\delta(r'-r_C) 
\nonumber \\ 
&=& -i \omega r_C^3 s^p_{lm\omega}
\left(h^{(1)}_l(\omega r)j_l(\omega r_C)\theta(r-r_C)
+j_l(\omega r)h^{(1)}_l(\omega r_C)\theta(r_C-r)\right) 
\,. 
\end{eqnarray}

The radial mode functions induced by the second, third, fifth and
sixth terms on the right hand side of (\ref{eq:sxx}) are obtained in a
closed form as
\begin{eqnarray}
F_{lm\omega}^{h(1)}(r) 
&=& \int dr' g_{lm\omega}(r,r') 
s^{h(1)}_{lm\omega}\left[
h^{(1)}_l(z)+{i \over z}{d \over dz} z h^{(1)}_l(z)
\right]_{z=\omega r'}\theta(r'-r_C)
\nonumber \\ 
&=& -{i\over \omega^2}s^{h(1)}_{lm\omega}\biggl[ 
\left\{
{i \over 4}z \left(h^{(1)}_{l+1}(z)-h^{(1)}_{l-1}(z)\right)
-{1 \over 2}(z-z_C)h^{(1)}_l(z)+f^{h(1)+}_{lm\omega}h^{(1)}_l(z)
\right\} \theta(r-r_C)
\nonumber \\ && \qquad\qquad\quad 
+f^{h(1)-}_{lm\omega}j_l(z)\theta(r_C-r)\biggr]
\,, \\ 
f^{h(1)+}_{lm\omega} 
&=& -{z_C^3 \over 4}\left(2j_l(z_C)h^{(1)}_l(z_C)
-j_{l-1}(z_C)h^{(1)}_{l+1}(z_C)-j_{l+1}(z_C)h^{(1)}_{l-1}(z_C)\right)
-{i \over 2}z_C^2j_l(z_C)h^{(1)}_l(z_C)
\,, \\ 
f^{h(1)-}_{lm\omega} 
&=& -{z_C^3 \over 2}\left(h^{(1)}_l{}^2(z_C)
-h^{(1)}_{l-1}(z_C)h^{(1)}_{l+1}(z_C)\right)
-{i \over 2}z_C^2h^{(1)}_l{}^2(z_C)
\,, \\ 
F_{lm\omega}^{h(2)}(r) 
&=& \int dr' g_{lm\omega}(r,r') 
s^{h(2)}_{lm\omega}\left[
h^{(1)}_l(z)+z{d \over dz} h^{(1)}_l(z)
\right]_{z=\omega r'}\theta(r'-r_C)
\nonumber \\ 
&=& -{i\over \omega^2}s^{h(2)}_{lm\omega}\biggl[ 
\left\{
-{i \over 8}z \left(h^{(1)}_{l+1}(z)-h^{(1)}_{l-1}(z)\right)
+{i \over 4}(z^2-z_C^2)h^{(1)}_l(z)+f^{h(2)+}_{lm\omega}h^{(1)}_l(z)
\right\} \theta(r-r_C)
\nonumber \\ && \qquad\qquad\quad 
+f^{h(2)-}_{lm\omega}j_l(z)\theta(r_C-r)\biggr]
\,, \\ 
f^{h(2)+}_{lm\omega} 
&=& -{z_C^3 \over 8}\left(2j_l(z_C)h^{(1)}_l(z_C)
+j_{l-1}(z_C)h^{(1)}_{l+1}(z_C)+j_{l+1}(z_C)h^{(1)}_{l-1}(z_C)\right)
\,, \\ 
f^{h(2)-}_{lm\omega} 
&=& -{z_C^3 \over 4}\left(h^{(1)}_l{}^2(z_C)
+h^{(1)}_{l-1}(z_C)h^{(1)}_{l+1}(z_C)\right)
\,, \\ 
F_{lm\omega}^{j(1)}(r) 
&=& \int dr' g_{lm\omega}(r,r') 
s^{j(1)}_{lm\omega}\left[
j_l(z)+{i \over z}{d \over dz} z j_l(z)
\right]_{z=\omega r'}\theta(r_C-r')
\nonumber \\ 
&=& -{i\over \omega^2}s^{j(1)}_{lm\omega}\biggl[ 
f^{j(1)+}_{lm\omega}h_l(z)\theta(r-r_C)
\nonumber \\ && \qquad\qquad
+\left\{
{i \over 4}z \left(j_{l+1}(z)-j_{l-1}(z)\right)
-{1 \over 2}(z-z_C)j_l(z)+f^{j(1)-}_{lm\omega}j_l(z)
\right\} \theta(r_C-r)\biggr]
\,, \\ 
f^{j(1)+}_{lm\omega} 
&=& {z_C^3 \over 2}\left(j_l^2(z_C)
-j_{l-1}(z_C)j_{l+1}(z_C)\right)
+{i \over 2}z_C^2j_l^2(z_C)
\,, \\ 
f^{j(1)-}_{lm\omega} 
&=& {z_C^3 \over 4}\left(2j_l(z_C)h^{(1)}_l(z_C)
-j_{l-1}(z_C)h^{(1)}_{l+1}(z_C)-j_{l+1}(z_C)h^{(1)}_{l-1}(z_C)\right)
+{i \over 2}z_C^2j_l(z_C)h^{(1)}_l(z_C)
\,, \\ 
F_{lm\omega}^{j(2)}(r) 
&=& \int dr' g_{lm\omega}(r,r') 
s^{j(2)}_{lm\omega}\left[j_l(z)
+z{d \over dz}j_l(z)\right]_{z=\omega r'}\theta(r_C-r')
\nonumber \\ 
&=& -{i\over \omega^2}s^{j(2)}_{lm\omega}\biggl[ 
f^{j(2)+}_{lm\omega}h_l(z)\theta(r-r_C)
\nonumber \\ && \qquad\qquad
+\left\{
-{i \over 8}z \left(j_{l+1}(z)-j_{l-1}(z)\right)
+{i \over 4}(z^2-z_C^2)j_l(z)+f^{j(2)-}_{lm\omega}j_l(z)
\right\} \theta(r_C-r)\biggr]
\,, \\ 
f^{j(2)+}_{lm\omega} 
&=& {z_C^3 \over 4}\left(j_l^2(z_C)
+j_{l-1}(z_C)j_{l+1}(z_C)\right)
\,, \\ 
f^{j(2)-}_{lm\omega} 
&=& {z_C^3 \over 8}\left(2j_l(z_C)h^{(1)}_l(z_C)
+j_{l-1}(z_C)h^{(1)}_{l+1}(z_C)+j_{l+1}(z_C)h^{(1)}_{l-1}(z_C)\right)
\,, 
\end{eqnarray}
where we use $z=\omega r$ and $z_C=\omega r_C$.

Unlike the previous terms, the radial mode functions 
induced by the fourth and seventh terms on the right hand side of (\ref{eq:sxx}) 
cannot be obtained in  closed form. 
For them, it is necessary to obtain 
the following indefinite integrals, 
\begin{eqnarray}
D^{(1)kk^*}_l = \int dz z k_l(z) k^*_l(z) \,, \quad 
D^{(2)kk^*}_l = \int dz z^3 k_l(z) k^*_l(z) \,, 
\end{eqnarray}
where $k$ and $k^*$ represent either $j$ or $h^{(1)}$. 
For finite $l$, these integrals can be obtained 
from the reccurence formula 
\begin{eqnarray}
D^{(1)kk^*}_l &=& D^{(1)kk^*}_{l-1} 
-{1 \over 2l}z^2\left(k_l(z)k^*_l(z)+k_{l-1}(z)k^*_{l-1}(z)\right)
\,, \\ 
D^{(2)kk^*}_l &=& {l+1 \over l-1}D^{(2)kk^*}_{l-1} 
-{1 \over 2(l-1)}z^4\left(k_l(z)k^*_l(z)+k_{l-1}(z)k^*_{l-1}(z)\right)
\,, 
\end{eqnarray}
with 
\begin{eqnarray}
D^{(1)jj}_{l=0} &=& {1 \over 2}\left(\ln(z)-{\rm Ci}(2z)\right) 
\,, \quad 
D^{(2)jj}_{l=0} = {z^2 \over 4}
-{z \over 4}\sin(2z)-{1 \over 8}\cos(2z)
\,, \\
D^{(2)jj}_{l=1} &=& {z^2 \over 4}
+{z \over 4}\sin(2z)+{5 \over 8}\cos(2z)
+{1 \over 2}\ln(z) -{1 \over 2}{\rm Ci}(2z)
\,, \\
D^{(1)jh}_{l=0} &=& {1 \over 2}\left({\rm Ei}(1,-2iz)+\ln(z)\right) 
\,, \quad 
D^{(2)jh}_{l=0} = {z^2 \over 4}+i{z \over 4}e^{i2z} 
-{1 \over 8}e^{i2z} 
\,, \\ 
D^{(2)jh}_{l=1} &=& {z^2 \over 4}-i{z \over 4}e^{i2z} 
+{5 \over 8}e^{i2z} +{1 \over 2}{\rm Ei}(1,-2iz) +{1 \over 2}\ln(z)
\,, \\
D^{(1)hh}_{l=0} &=& {\rm Ei}(1,-2iz)
\,, \quad 
D^{(2)hh}_{l=0} = i{1 \over 2}z e^{2iz}-{1 \over 4}e^{2iz}
\,, \\ 
D^{(2)hh}_{l=1} &=& -i{1 \over 2}z e^{2iz}+{5 \over 4}e^{2iz}
+{\rm Ei}(1,-2iz)
\,, 
\end{eqnarray}
where ${\rm Ci}(z) = -\int^\infty_z dz \cos(z)/z$ 
is the cosine integral function 
and ${\rm Ei}(1,z) = -\int^\infty_1 dz e^{-zt}/t$ 
in the first exponential integral function.

Using these integrals, we have 
\begin{eqnarray}
F_{lm\omega}^{h(3)}(r) 
&=& \int dr' g_{lm\omega}(r,r') 
s^{h(3)}_{lm\omega}\left[-i2z h^{(1)}_l(z)
+i {l(l+1) \over z}h^{(1)}_l(z)\right]_{z=\omega r'}\theta(r'-r_C) 
\nonumber \\ 
&=& -{i\over \omega^2}s^{h(3)}_{lm\omega}\Biggl[
\biggl\{h^{(1)}_l(z) D^{(2)jh}_l(z) 
-j_l(z) D^{(2)hh}_l(z)-h^{(1)}_l(z) D^{(2)jh}_l(z_C)
\biggr\} \theta(r-r_C)
\nonumber \\ && \qquad\qquad\quad 
-j_l(z) D^{(2)hh}_l(z_C)\theta(z_C-z)\Biggr]
\,, \\ 
F_{lm\omega}^{j(3)}(r) 
&=& \int dr' g_{lm\omega}(r,r') 
s^{j(3)}_{lm\omega}\left[-i2z j_l(z)
+i {l(l+1) \over z}j_l(z)\right]_{z=\omega r'}\theta(r_C-r') 
\nonumber \\ 
&=& -{i\over \omega^2}s^{j(3)}_{lm\omega}\Biggl[ 
h^{(1)}_l(z) D^{jj}_l(z_C)\theta(z-z_C)
\nonumber \\ && \qquad\qquad\quad 
+\biggl\{h^{(1)}_l(z) D^{(2)jj}_l(z) 
-j_l(z) D^{(2)jh}_l(z)+j_l(z) D^{(2)jh}_l(z_C)
\biggr\} \theta(r-r_C)\Biggr]
\,, 
\end{eqnarray}
where we use 
$D^{kk^*}_l = -i2D^{(2)kk^*}_l+il(l+1)D^{(1)kk^*}_l$.



\end{document}